\documentclass[aps,twocolumn,amsmath,amssymb,showpacs,prl,unsortedaddress]{revtex4-1} 
\usepackage{epsf}      
\usepackage{graphicx}
\usepackage{gensymb}
\usepackage{sidecap}

\begin{document}

\title {What controls the phase diagram and superconductivity in Ru substituted BaFe$_2$As$_2$?}

\author{R. S. Dhaka}
\email{rsdhaka@ameslab.gov}
\author{Chang Liu}
\author{R. M. Fernandes}
\author{Rui Jiang}
\author{C. P. Strehlow}
\author{Takeshi Kondo}
\author{A. Thaler}
\author{J\text{\"o}rg Schmalian}
\author{S. L. Bud'ko}
\author{P. C. Canfield}
\author{Adam Kaminski}
\email{kaminski@ameslab.gov}

\affiliation{Division of Materials Science and Engineering, 
The Ames Laboratory, U.S. DOE and Department of Physics and Astronomy, Iowa State University, Ames, Iowa 50011, USA}

\date{\today}                                         

\begin{abstract}
We use high resolution angle-resolved photoemission to study the electronic structure of the iron based high-temperature superconductors Ba(Fe$_{1-x}$Ru$_x$)$_2$As$_2$ as a function of Ru concentration. We find that substitution of Ru for Fe is isoelectronic, i. e., it does not change the value of the chemical potential. More interestingly, there are no measured, significant changes in the shape of the Fermi surface or in the Fermi velocity over a wide range of substitution levels ($0<x<0.55$). Given that the suppression of the antiferromagnetic and structural phase has been associated with the emergence of the superconducting state, Ru substitution must achieve this via a mechanism that does not involve changes of the Fermi surface. We speculate that this mechanism relies on magnetic dilution which leads to the reduction of the effective Stoner enhancement. 
\end{abstract}

\pacs{74.25.Jb,74.62.Dh,74.70.-b,79.60.-i}
\maketitle

External control parameters such as pressure or chemical substitution play an important role in extending the phase space of novel materials with interesting and useful properties. An excellent example is the case of the FeAs family of antiferromagnets, that can be turned into high temperature superconductors by substituting with transition metals or application of pressure\cite{Kamihara08,Rotter08,SefatPRL08,CanfieldRev10,SasmalPRL08,Colombier09,Kimber09,Johnston10}. The close relationship between the superconducting (SC) and the antiferromagnetic (AFM) state suggest the presence of an electronic pairing mechanism\cite{VorontsovPRB09,FernandesPRB10,ParkerPRB09,WangPRL09}. In some materials, these two competing orders even coexist microscopically \cite{CanfieldRev10,PrattPRL09,DrewNM09}. In fact, when doping is introduced either inside or in between the FeAs planes, SC only develops after the AFM transition temperature ($T_{\rm N}$) is sufficiently suppressed, and the highest values of $T_{\rm c}$ are achieved close to the concentration where the AFM state ceased to exist\cite{CanfieldRev10}. Therefore, the investigation of the correlation between AFM and SC is essential to the understanding of the microscopic pairing mechanism. Although the details of this remarkable transition are not fully understood, there is empirical evidence that chemical substitution of atoms of one element in the crystal by a different element affects the electronic structure in two different ways. It can change the value of the chemical potential ($\mu$) if the substitution element adds charge carriers. For example, in Ba(Fe$_{1-x}$Co$_x$)$_2$As$_2$, Co adds electrons and increases $\mu$. The resulting changes of the Fermi surface size and even topology (Lifshitz transitions)\cite{Mun09,LiuPRBNP} are empirically associated with the onset and offset of the SC state. An interesting question was raised based on theoretical calculation suggesting that substitution of Fe by Co(Ni) in iron arsenides does not change carrier number, instead the extra electrons are localized around the impurity atoms and such substitution should effectively be isoelectronic \cite{WadatiPRL10}. Suprisingly, the Fermi surface, band dispersion and the total number of extra carriers are experimentally shown to change with increasing substitution at least in case of Co \cite{LiuPRBNP,Konbu11,Aswartham11,Neupane11}. Nevertheless, the importance of impurity scattering in destabilizing the spin density wave state and enabling superconductivity is likely a very important aspect of the physics of these materials \cite{WadatiPRL10,Konbu11,Nakamura11}.

\begin{figure*}
\includegraphics[width=7.25in]{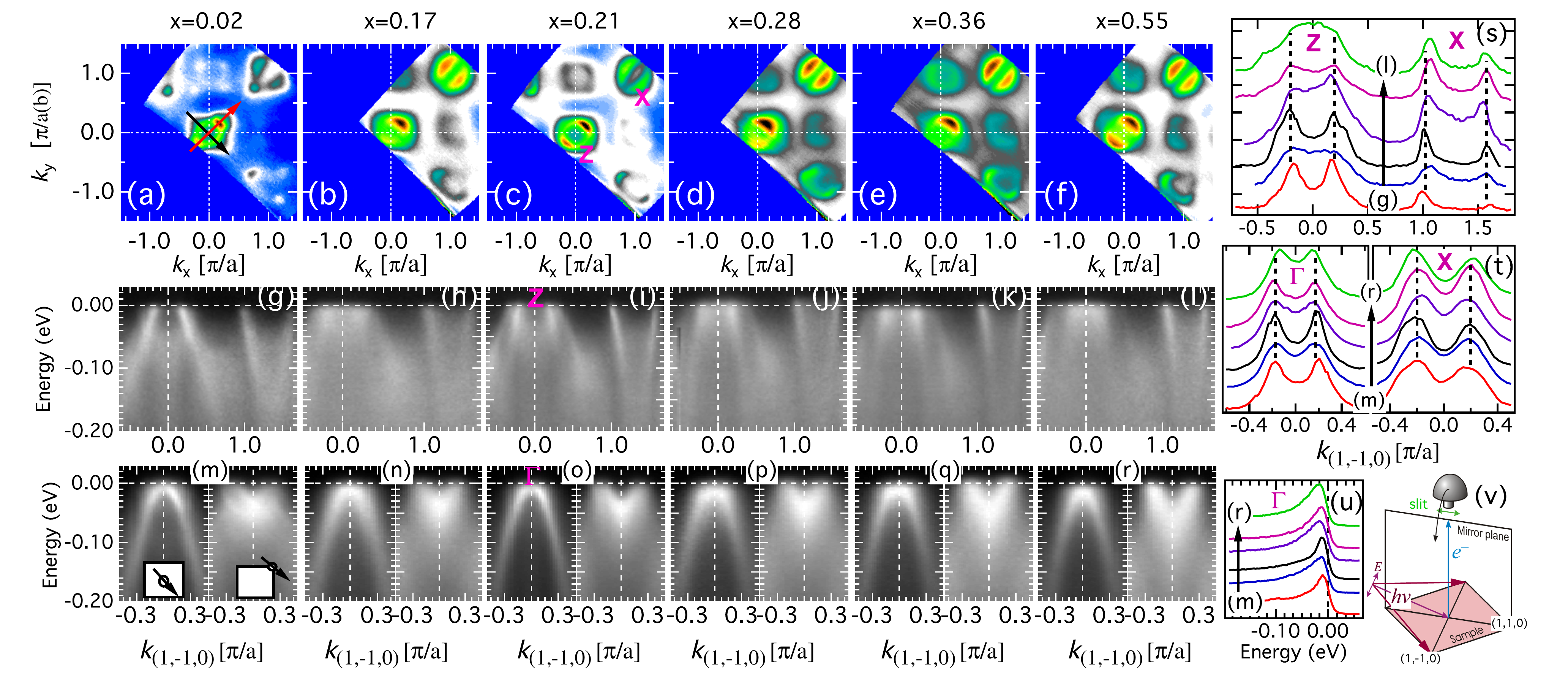}
\caption{(color online) (a-f) FS maps (around the upper zone edge $Z$) of Ba(Fe$_{1-x}$Ru$_x$)$_2$As$_2$ and the band dispersion data (along the $Z-X$ direction) measured with h$\nu=35$~eV (g-l) and 49~eV (m-r) for various $x$. The photoemission intensity map is obtained by integrating over an energy window of $E_{\rm F}$$\pm$10~meV, white areas mark high photoelectron intensity and thus the locations of the bands. The black arrow, shown in (a), indicates the cutting direction of the band dispersion plots in (g-l). The same direction is used to extract the pocket size and throughout the paper until unless not defined.  MDCs are shown in (s)  at $E_{\rm F}$ for (g-l) and in (t) left panel at 50~meV below $E_{\rm F}$ and right panel at $E_{\rm F}$ for (m-r). (u) show the EDCs from left panel of (m-r). Red arrow in (a) is used for the data (plotted in red) in Fig.~3d and for the band dispersion in Fig.~4a (red arrow marked in inset). (v) Schematics of the experimental setup where the electric-field vector of the incoming light is polarized along the $k_{\rm (1,-1,0)}$  direction and the entrance slit of the electron analyzer is along the mirror plane.}
\label{fig1}
\end{figure*}

On the other hand, substitution by an isoelectronic element ($\it e. g.$ As with P\cite{Kasahara10,JiangKlintberg}) can change the lattice constants in a similar way as application of external pressure\cite{Colombier09,Kimber09,Zhang09}, which is known to induce superconductivity. This is thought to modify the bandwidth and hybridization which leads to a change in the shape of the FS \cite{Thirupathaiah11,ShishidoPRL10}, while preserving the carrier concentration, however, another ARPES study show that P doping effectively induce holes into the system\cite{FengP11}. An interesting case is the substitution of Fe by Ru\cite{Thaler10,SharmaSchnelle}, where the mechanism of suppressing the AFM and inducing superconductivity is less obvious, as the ionic radii of the Ru is larger than that of Fe. Detailed studies in Ref.~\onlinecite{Thaler10} reveal that the changes in the lattice constants are quite intricate with the $c$ axis lattice constant shrinking and the in-plane lattice constants expanding. The overall unit cell volume increases\cite{Thaler10}, in contrast to P substitution\cite{Kasahara10,JiangKlintberg}. At first sight, one expects Ru substitution to affect the bandwidth and hybridization, but not the chemical potential. Indeed, it has been predicted that the isoelectronic Ru substitution in oxypnictides does not change the carrier concentration as well as the electronic structure\cite{Nakamura11,Tropeano10}. However, it is also possible that the dopant assumes different valence states, specially in the case of the $4d$ element Ru, which would effectively introduce carriers in the system and change the FS\cite{SharmaSchnelle,ZhangWang}. Moreover, the band structure of Ba(Fe$_{1-x}$Ru$_x$)$_2$As$_2$ by using density functional theory suggest that Ru substitution does not increase the number of carriers but does increase the broadening of the $d$ bands via hybridization\cite{ZhangWang}. Recent ARPES measurements suggested that there are significant differences between the FS of Ba(Fe$_{0.65}$Ru$_{0.35}$)$_2$As$_2$ and of the parent compound\cite{BrouetPRL10}. However this study compared the band dispersion at high temperature in the paramagnetic state of the parent compound with one measured at low temperature for the Ru substituted sample. Due to the thermal expansion, the lattice constants \cite{Budko09,Luz09} and band structure in pnictides change significantly with temperature\cite{LiuPRL09,Dhaka}. Therefore, it is important to compare the FS with various $x$, measured at the same  temperatures. 

In this Letter we demonstrate that the chemical potential and FS shape of Ba(Fe$_{1-x}$Ru$_x$)$_2$As$_2$ does not change significantly for a wide range of Ru concentration ($0<x<0.55$). Thus the substitution of Fe with isoelectronic Ru seems to be unique, since it does not change the low energy electronic excitation spectrum, yet it results in a similar phase diagram including a superconducting   dome. The most likely explanation of our findings is that magnetic dilution and the associated reduction of the effective Stoner enhancement or impurity scattering leads to the suppression of the AFM order.  It is quite remarkable that the mere suppression of the AFM order, regardless of the way in which it is achieved, is necessary for establishing the superconductivity in this class of materials. It is equally remarkable that superconductivity is robust even after 40\% of Fe atoms are replaced by Ru. 

Single crystals of Ba(Fe$_{1-x}$Ru$_x$)$_2$As$_2$ were grown out of self-flux using conventional high-temperature solution growth techniques and studied by the transport and magnetization measurements\cite{Thaler10,Hodovanets11}. The ARPES measurements were performed (in grazing incidence geometry) at beamline 10.0.1 of the Advanced Light Source (ALS), Berkeley, California using a Scienta R4000 electron analyzer. The schematics of the experimental geometry is shown in Fig.~1(v). Core level data were taken with h$\nu$=60~eV at the PGM beamline of the Synchrotron Radiation Center (SRC), Wisconsin. The measurements at Ames Laboratory were acquired using a Scienta SES2002 electron analyzer and a GammaData  helium ultraviolet lamp. The samples were mounted on an Al pin using UHV compatible epoxy and {\it in situ} cleaved  perpendicular to the $c$-axis, yielding single layer surfaces in the $a$-$b$ plane. All ARPES data (except Fig.~2) were collected at sample temperature of $\approx$15~K and in ultrahigh vacuum below 4$\times$10$^{-11}$ torr. The energy and angular resolutions were set at 15~meV and ~0.3$^{\circ}$, respectively. High symmetry points were defined the same way as in Ref.~\onlinecite{LiuPRBNP}. Measurements carried out on several samples yielded similar results for the band dispersion and FS.

\begin{figure}
\includegraphics[width=3.55in]{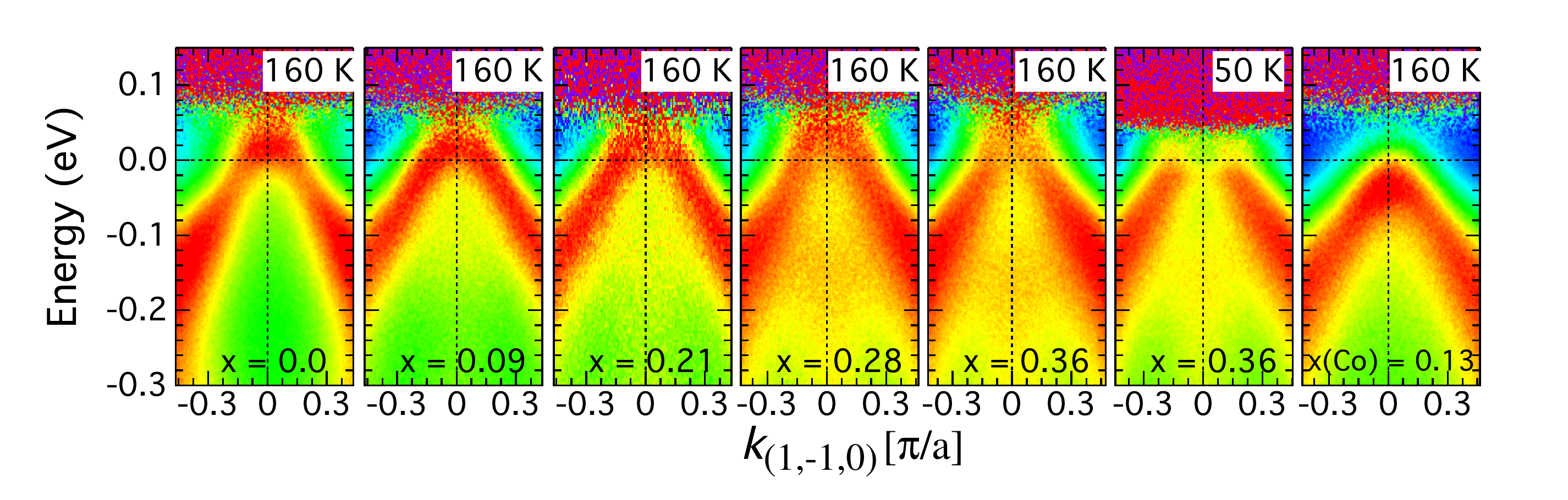}
\caption{(color online) The band dispersion data measured with He~II photon source (h$\nu=40.8$~eV, $k_z\simeq1.1\pi/c$).}
\label{fig2}
\end{figure}

Figures~1(a-f) show the FS topology for different Ru concentration, measured at h$\nu$ = 35~eV ($k_z\simeq 2\pi/c$) {\it i.e.} the upper edge of the first Brillouin zone ($Z$ point)\cite{KondoPRB10}. Two almost degenerate $\alpha$ and $\beta$ hole pockets centered at the $Z$ ($\Gamma$) (0,0) point and two electron pockets centered at the $X$ point have been observed in the 2D Brillouin zone, as predicted from band structure calculations. The feature between adjacent $X$ points was  identified as arising due to surface reconstruction\cite{Hsieh08}. In Figs.~1(g-l), we plot the corresponding band dispersion data (h$\nu$ = 35~eV) along the direction marked by the black arrow in Fig.~1(a).  Figures~1(m-r) show the band dispersion measured with h$\nu$ = 49~eV ($k_z\simeq 0$) and plotted along the directions shown in Fig.~1(m). The low energy band dispersion and FS for all Ru concentrations shown here are very similar to that of the parent BaFe$_2$As$_2$ compound\cite{KondoPRB10}. We also compared the band dispersion data (divided by the Fermi function) in Fig.~2. Note that the measurements at 160~K and 50~K for the $x=0.36$ (Fig.~2), demonstrate that the band structure in these materials strongly depends on temperature, in agreement with previous reports\cite{ LiuPRL09}.  This dependence of the band structure on temperature is likely the explanation of disagreement between results presented here and in Ref.~\onlinecite{BrouetPRL10}. We contrast the lack of effect of Ru substitution on low energy band structure with significant shift of the chemical potential by Co substitution in last panel of Fig. 2, where the hole pockets vanishes already at $x=0.13$. It is quite striking that both FS maps and band dispersion data do not visibly change when Ru concentration is increased from 2\% to 55\% - a range of substitution much larger than the span of the SC dome\cite{Thaler10} shown in Fig.~3(a). However, the substitution of Fe with Ru clearly has to have some effect on the electronic structure. Indeed, the valence band spectra reflect the density of states for each sample [shown in Fig.~3(b)] and show the presence of peaks characteristic of elemental Ru (inset), with their intensity increasing with Ru concentration. Note that the energy position of the features will be slightly different depending whether it is metallic Ru, or Ru ions chemically bound to other atoms. We simply mark the features in both spectra that originate from Ru orbitals.

\begin{figure}
\includegraphics[width=3.6in]{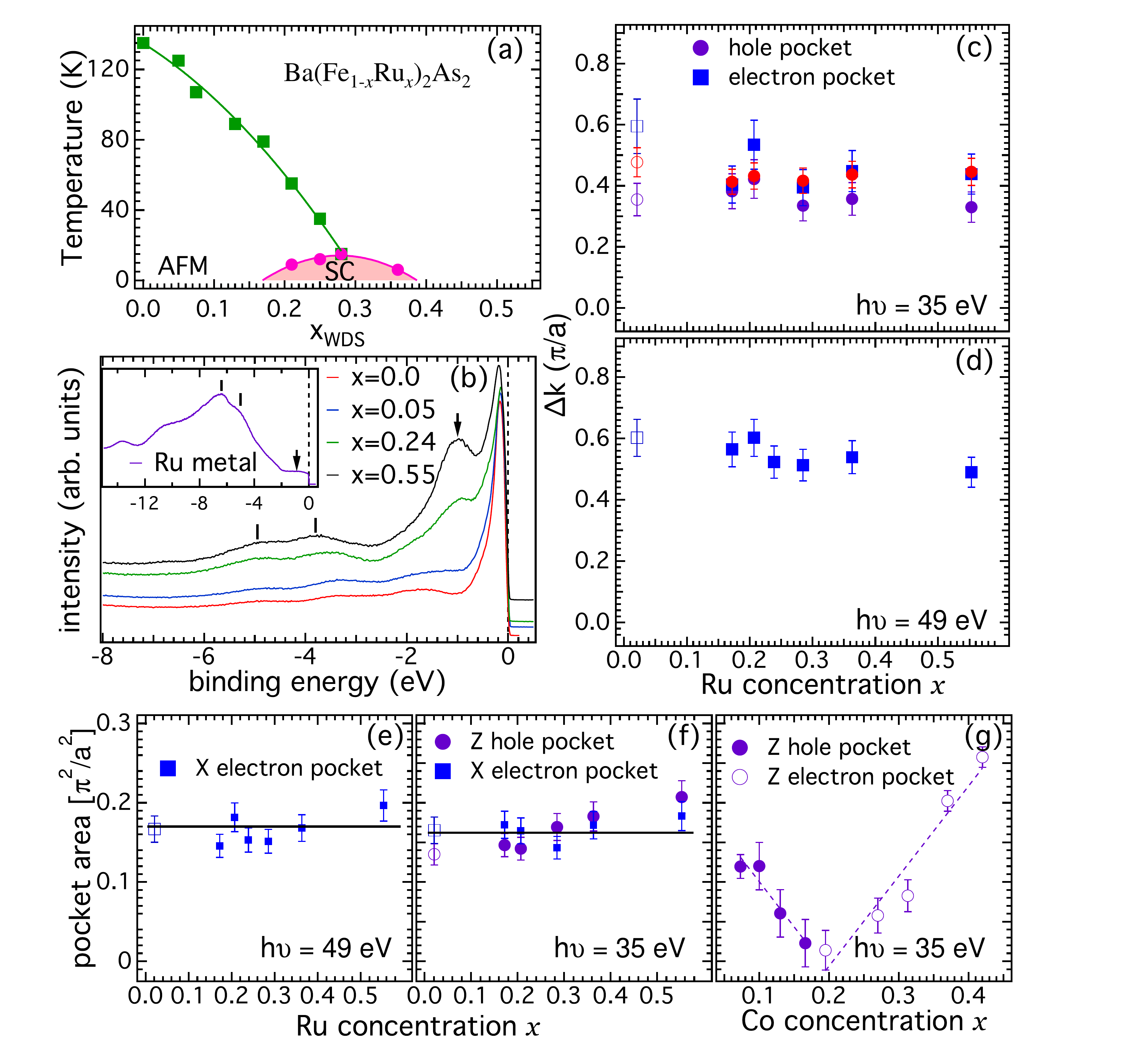}
\caption{(color online) (a) Schematic phase diagram of Ba(Fe$_{1-x}$Ru$_x$)$_2$As$_2$. (b) The shallow core-level spectra (h$\nu=60$~eV) for various $x$ (inset shows for polycrystalline Ru metal). These spectra have been normalized to the same height in the same way and offset is used along the vertical axis for clarity of presentation. $\Delta$$k_{\rm F}$ for hole (electron) pocket measured with (c) 49~eV ($k_z=0$) and (d) 35~eV ($k_z\simeq 2\pi/c$). Open symbols indicate data measured in AFM state. In (d), violet color is used for $\alpha$ band and red is used for $\beta$ band [see arrows in Fig.~1(a)]. (e) the plot of $\Gamma$ and $X$ pocket area. (f) the $Z$ and $X$ pocket area. (g) the $Z$ pocket area with Co concentration is shown for comparison adopted from Ref.~\onlinecite{LiuPRBNP}.}
\label{fig3}
\end{figure}

In order to quantify the evolution of the size of both hole and electron pockets with concentration, we extracted values of the Fermi momenta $k_{\rm F}$ from peaks in the momentum distribution curves (MDCs) at  $\mu$, associated with data in Fig.~1 as well as data obtained in the center of the BZ (encompassing the $\Gamma$ point). The resulting sizes of the Fermi surface $\Delta k_{\rm F}$ are shown in Figs.~3(c,d) and the corresponding areas of the pockets (averaged over $\alpha$ and $\beta$ bands) are shown in Figs.~3(e,f). Although the samples below $x\approx0.28$ are in the AFM state [Fig.~3(a)], we note that only low substitution level, $x=0.02$ sample, (open symbols in Figs.~3-5) shows reconstructed FS (Fig.~1)\cite{LiuPRBNP}. Both the Fermi momentum and total area of all FS pockets remains surprisingly constant over the range of Ru concentration where the SC dome exists in the phase diagram. However, slight change can not be ruled out for $x\ge0.28$ which might be related to the transition from AFM to paramagnetic phase. This clearly demonstrated that Ru substitution is isoelectronic and preserves the carrier concentration at least up to 40\% substitution. We contrast this unusual behavior with Ba(Fe$_{1-x}$Co$_x$)$_2$As$_2$, where even small substitutions induce large changes not only in the size of the FS pockets [Figs.~2 and 3(g)] but also in the FS topology\cite{LiuPRBNP}. In that case the superconductivity emerges and vanishes close to the doping levels where the FS topology changes, i. e., Lifshitz transition\cite{LiuPRBNP}. Here, we demonstrate that Ru substitution controls the phase diagram in a very different way. It does not change the paramagnetic band structure\cite{Nakamura11}  nor the chemical potential. The changes between $x=0.02$ and $x=0.17$ (Fig.~1) originate from the reduction of the magnetically reconstructed FS to its paramagnetic appearance where band back folding no longer occurs. Surprisingly, the low temperature transition from AFM to PM state occurs without changing the FS nesting condition (Fig.~3).

\begin{figure}
\includegraphics[width=3.5in]{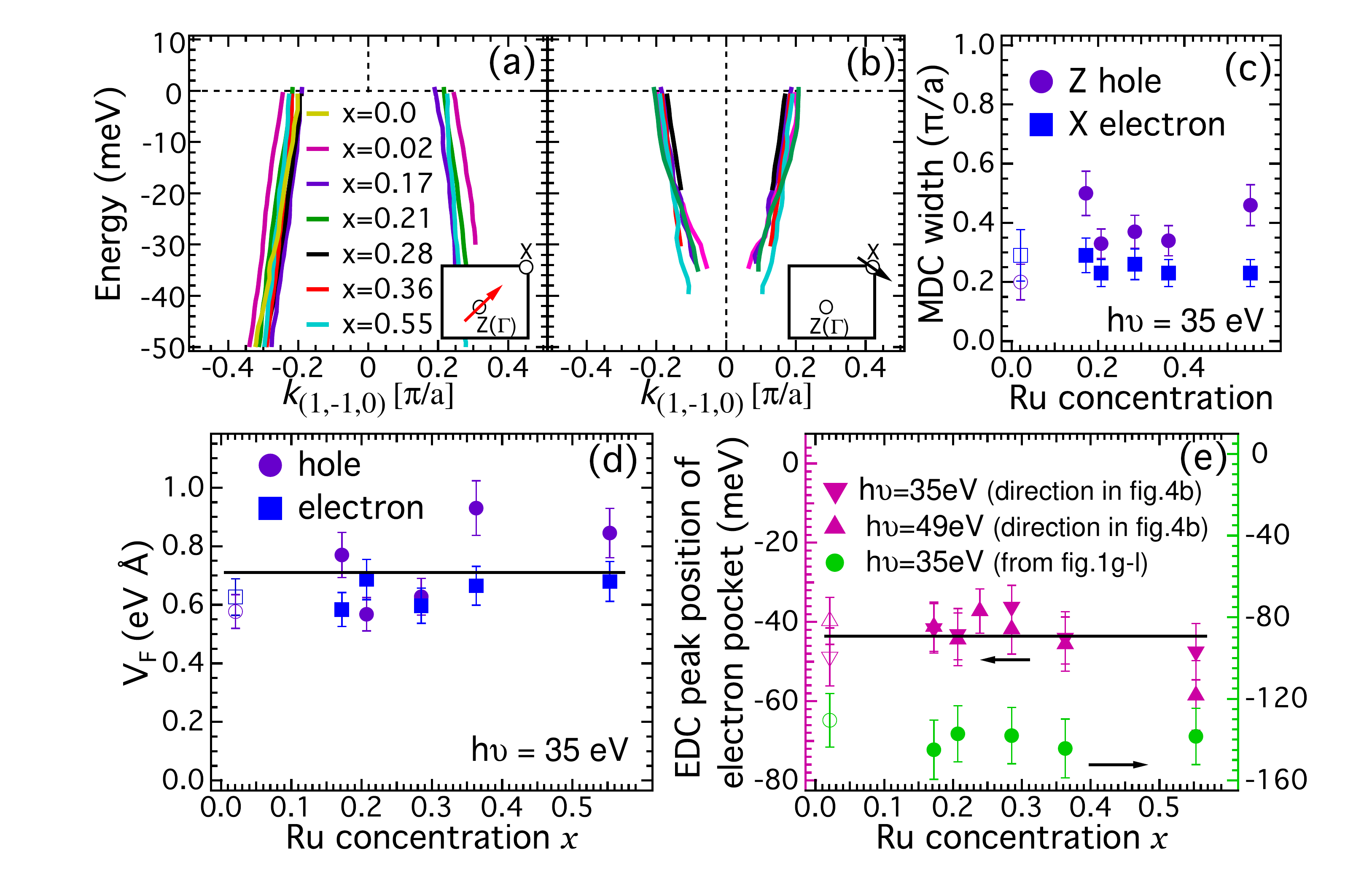}
\caption{(color online) (a-b) The band dispersion obtained by MDC fitting (averaged over $\alpha$ and $\beta$ bands) along the direction shown in the inset and (c) MDC width (h$\nu$ = 35~eV). (d) Fermi velocity extracted from data in (a,b) near $E_{\rm F}$. (e) location of the bottom of the electron band extracted from Energy Distribution Curves.}
\label{fig4}
\end{figure}

The photoemission data in Fig.~3(b) demonstrates the appearance and shift in energy of valence band peaks that are mostly confined to higher binding energies. This is a result of the introduction of foreign orbitals (in this case Ru) to the sample. Typically, the external or chemical pressure modifies both the bandwidth and hybridization, which leads to reshaping of the FS and significant changes in the Fermi velocity\cite{Zhang09,Thirupathaiah11}. We now examine the changes in details of the low energy band dispersion, MDC width and Fermi velocity ($V_{\rm F}$) upon Ru substitution. In Figs.~4(a,b) we plot the band dispersion extracted by fitting MDCs with Lorentzian peaks. The  MDC width and Fermi velocities are plotted in Figs.~4(c) and 4(d), respectively. Despite some sample to sample variation, neither the band dispersion and MDC width nor the $V_{\rm F}$ change significantly in a systematic way with $x$. The main source of error bars in the dispersion are likely the sample flatness and positioning.This is indeed very remarkable, since the suppression of AFM order occurs here without changes in the FS, unlike in case the of P or Co substitution \cite{LiuPRBNP,Thirupathaiah11,FengP11}. The EDC peak position of the bottom of the electron pocket also does not show significant change with $x$ [Fig.~4(e)]. It appears that any bandwidth changes due to introduction of Ru are confined to higher binding energies and do not affect low energy electronic excitations. This is quantitatively in agreement with the recent band structure calculation where it is shown that unlike Co (Ni), Ru substitution does not affect the low energy band dispersions of the iron arsenide\cite{Nakamura11}.

Having established that no significant change occurs in the FS with $x$ at two $k_z$ points, it is important to check if this holds for all values of $k_z$, since these materials have 3D electronic structure\cite{LiuPRL09, YoshidaPRL11,VilmercatiPRB09,ThirupathaiahPRB10}. The data in Fig.~5 reveals that indeed the band dispersion along $k_z$ does not vary with Ru substitution. To make more quantitative comparison, we fitted the MDC peaks and extracted the pocket size for various $k_z$ points [Fig.~5(e)].  Indeed, no changes were observed within reasonable experimental error bars. Note that, the electron pockets do not exhibit strong $k_z$ dispersion (not shown).

\begin{figure}
\includegraphics[width=3.55in]{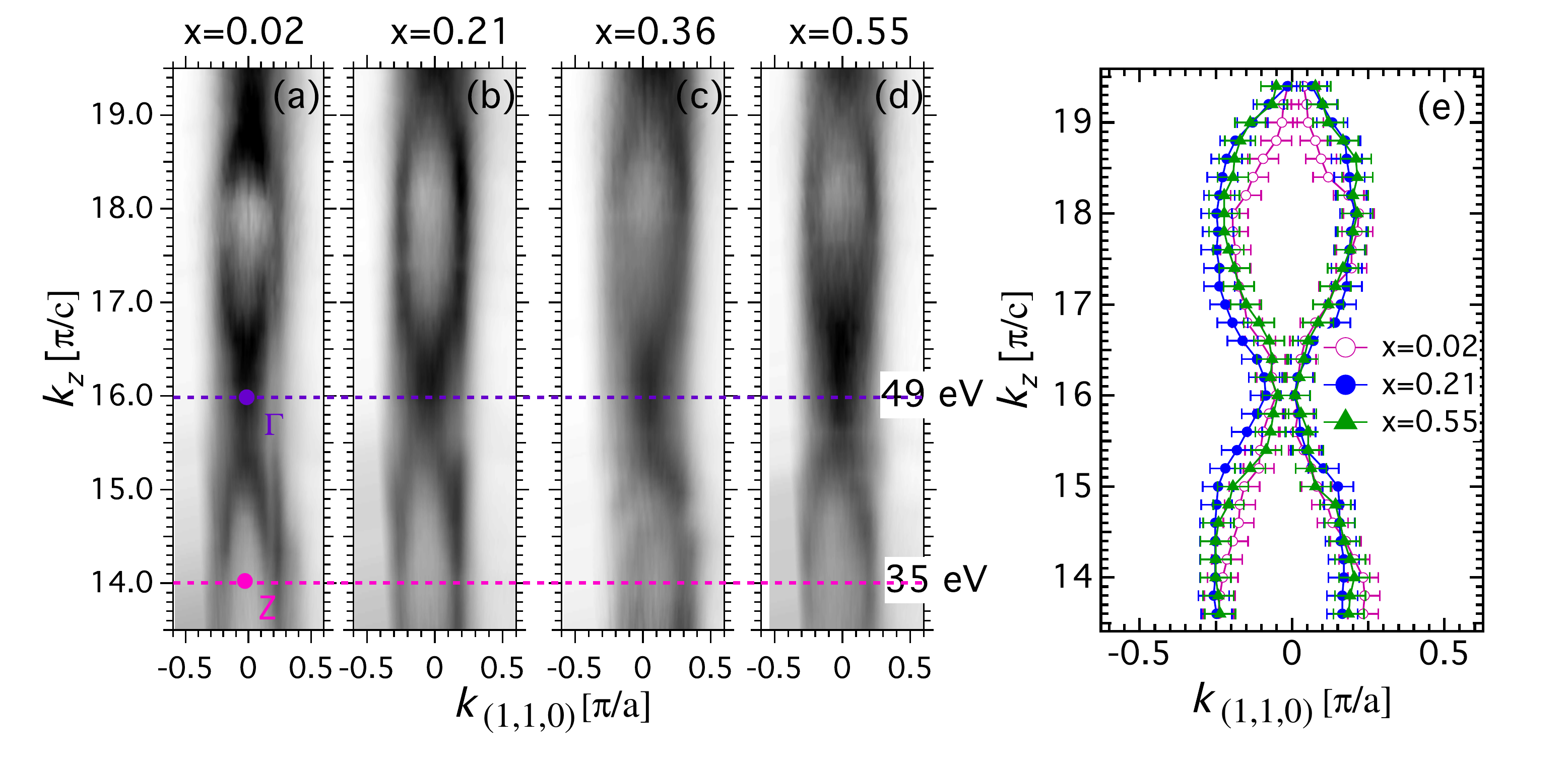}
\caption{(color online) (a-d) the hole FS maps along the $k_{||}$-$k_z$  plane measured around the center of the BZ by changing h$\nu$. (e) the fitted MDC peaks (averaged over $\alpha$ and $\beta$ bands) for various $x$ as a function of $k_z$.}
\label{fig5}
\end{figure}

In conclusion, we demonstrate that the chemical potential and FS shape in Ba(Fe$_{1-x}$Ru$_x$)$_2$As$_2$ does not change significantly for wide range of Ru concentration ($0<x<0.55$). These unexpected results suggest that Ru substitution tunes the properties of the FeAs superconductors in a different way than carrier doping or pressure. Instead of de-tuning the nesting condition of the Fermi surface to weaken the AFM order, Ru substitution seems to act in similar fashion to magnetic dilution. In the parent compound, the magnetic instability is a consequence not only of the nesting properties of the Fermi surface, but also of the proximity between the values of the Stoner enhancement parameter ($I$) and the inverse of the density of states at the Fermi level, $1/N(E_{\rm F})$. Being a $4d$ element, Ru has a smaller $I$ than the $3d$ element Fe, since $4d$ orbitals are much more extended than $3d$ orbitals. Therefore, introduction of sufficient number of incompatible orbitals upon Ru substitution reduces effective $I$. On the other hand, our data on the $x$ dependence of $\Delta k_{\rm F}$ (Fig.~3), MDC width  and $V_{\rm F}$ (Fig.~4) indicate that the density of states $N(E_{\rm F})$ remains practically unchanged. Thus, although the nesting properties of the FS are the same across the phase diagram, the reduction of $I$ upon magnetic dilution can lead to the suppression of both the magnetic transition temperature and the static magnetic moment, as observed by neutron diffraction\cite{Kim11}. Band structure calculations address a similar decrease in $I$ with Ru substitution, although they also predict small changes in $N(E_{\rm F})$ and in the Fermi surface topology, particularly in the $k_z$ dispersion of the central hole pockets \cite{ZhangWang} which are not supported by our measurements. It is also possible that Ru also acts as an impurity scatterer, reducing $T_{\rm N}$ and $T_{\rm c}$. In this regard, it is remarkable that the maximum $T_{\rm c}$ is fairly close to the value observed in optimally Co doped samples, even though the amount of disorder introduced by Ru in the Fe plane is rather large ($\approx30\%$). Thus, it remains a challenge to develop a complete microscopic model which is able to account for magnetic dilution, impurity scattering and robust superconductivity.

We thank Sung-Kwan Mo for their excellent instrumental support at the ALS. The Ames Laboratory is supported by the DOE - Basic Energy Sciences under Contract No. DE-AC02-07CH11358. The Advanced Light Source is supported by the Director, Office of Science, Office of Basic Energy Sciences, of the U.S. Department of Energy under Contract No. DE-AC02-05CH11231. SRC is supported by National Science Foundation under Award No. DMR-0537588.

\end{document}